          [2001/04/25 1.1 (PWD)]
\documentclass[a4paper,twocolumn]{esapub2005} 
\usepackage{times}
\usepackage{natbib}
\usepackage{graphicx}
\usepackage{latexsym}
\usepackage{amsmath}
\usepackage{amsfonts}
\usepackage{amssymb}

\title{Testing the Metric-Field Equations of Gravitation}
\author{Leonid V. Verozub} 
\affil{Kharkov National University, Kharkov 610077  , Ukraine}

\begin{document}

\keywords{gravitation; compact objects; cosmology; binary pulsar }

\maketitle
\begin{abstract}
Physical consequences from gravitation equations  based on Poincar\'{e} ideas of relativity of space and time in respect  of
measuring instruments  are considered. The most interesting  of them are  the possibility of the existence of stable 
supermassive configurations $(10^{2} \div 10^{12}\, M_{\odot})$  which can exist in galactic centres, and an explanation of the
acceleration of the Universe  expansion as a manifestation of the  gravitational force  properties in the theory under consideration.
\end{abstract}

\section{Introduction}

At the beginning of the 20th century H. Poincar\'{e}
convincingly showed that there is no
   sense to assert that one
or other geometry of physical space is a true. Only the aggregate
 " geometry + measuring instruments" has a physical, verifiable  by
experience sense. A. Einstein recognised the justice  of
Poincar\'{e}'s reasons. His theory of gravity does not reject
Poincar\'{e}'s ideas. It only demonstrates  relativity of space and time with respect to
distribution of matter.
Poincar\'{e}'s
ideas indicate to another property of physical reality - relativity
of the  geometrical properties with respect to  measuring
instruments, dependency on their properties. These ideas,
apparently, have never been realized in physics and remain until
now, more likely, by
 a subject for  discussions of philosophers. The success  of
General Relativity almost convinced us  that  physical space-time
in the presence of matter is Riemannian and that this fact does
not depend on properties of the measuring instruments. The
attempts to describe gravity in flat space-time did not obtain
recognition and Poincar\'{e}'s ideas proved to be almost forgotten
for physics.

However, are Poincar\'{e}'s ideas really only a thing of such little use as
conventionalism, as it is usually believed? A choice of certain properties
of measuring instruments is nothing but the choice of some frame of
reference, which is just such a physical device by means of which we test
properties of space-time. (Of course, we must understand a distinction
between a frame of reference as a physical device, and a coordinate system
as only a means of parameterisation of events in space-time). For this
reason Poincar\'{e}'s reasoning about interdependency between properties of
space-time and measuring instruments should be understood as the existence
of a connection between geometry of space-time and properties of the  frame
of reference being used. \citep{verozub01,  verozub02}

The existence of such a connection can be shown by an analysis of a simple
and well-known example. 
Disregarding the rotation of the Earth we can consider a reference
frame, rigidly connected with the Earth surface, as the inertial
one (IFR). An observer, who is located in this frame of reference,
can, without coming into conflict with  experience, describe the
motion of a test particle as the one  in Minkowski  space-time,
under the action of a certain force field $\mathcal{F}$. Consider now an
observer, who is located in a frame of reference, the reference
body of which formed by identical dot masses , free falling in the
field $\mathcal{F}$. Such a reference frame can be named the proper frame of
reference (PFR) for the given field. Assume,  the observer
is deprived of the possibility to see the Earth and stars. This
observer does not feel the presence of force field $\mathcal{F}$ in any
place of the frame. Therefore, if he proceeds from  relativity
of space-time in Berkeley-Leibnitz-Mach- Poincar\'{e} (BLMP)
meaning, then from his viewpoint the accelerations of the dot masses,
forming the reference body of his  frame, in his physical space
must be equal to zero. However, instead, he observes a change in
the distances between these dot masses in time. How  can he
explain this fact? Evidently,  only reasonable explanation for him
is the interpretation of this observed phenomenon as a
manifestation of the deviation of geodetic lines in a Riemannian
space of nonzero curvature. 

 Thus, if an observer in the IFR postulates that
space-time is flat, then the observer in a proper reference
frames of the force field $\mathcal{F}$, who proceeds from  relativity of
 space and time in the BLMP meaning, \textit{is forced} to
consider it as Riemannian.

\section{Metric-Field Equations of gravitation}
We accept according to the Special Relativity that space-time in
IFRs is Pseudo-Euclidean.  However can be argued  \citep{verozub01, verozub02}  that
 if we proceed  from relativity of space-time in the BLMP sense,
 then the differential metric form of space-time in  PFRs
is of the form 
\begin{equation}  \label{Vr_5/4}
ds = -(m_{p} c)^{-1} \; dS(x,dx).
\end{equation}
In this equation  $dS=\mathcal{L}(x,\dot{x})dt$  and $\mathcal{L}(x,\dot{x})$ is
the Lagrange function  describing in an
IFR  the motion of  identical dot  masses $m_{p}$, forming the
 reference body of the PFR. Thus, the properties of the space-time in a
PFR are defined by  properties of the used frame
in accordance with the   Poincar\'{e}  idea.

In order to describe gravitation in inertial and proper frames of reference  
we suppose   \citep{thirring}
that in pseudo- Euclidean space-time gravitation can be described
as a tensor field  $\psi_{\alpha\beta}(x)$ of spin 2, and the Lagrangian,
describing the motion of a test particle with the mass $m_{p}$, is given by
the form
\begin{equation}
\mathcal{L}=-m_{p} c [g_{\alpha\beta}(\psi)\;\dot{x}^{\alpha}\;\dot
{x}^{\beta}]^{1/2},  \label{LagrangianThirr}
\end{equation}
where $\dot{x}^{\alpha}=dx^{\alpha}/dt$ and $g_{\alpha\beta}$ is a
symmetric tensor whose components are 
functions of $ \psi_{\alpha\beta} $.

If particles move under influence of the force field $\psi_{\alpha\beta}(x)$,  then according to (\ref{Vr_5/4})
 the space-time metric differential form in PFRs of this field takes the form
\begin{equation}
ds^{2}=g_{\alpha\beta}(\psi)\;dx^{\alpha}\;dx^{\beta}
\end{equation}
Therefore,  the space-time in
the PFRs is  Riemannian    with  curvature other than zero.
The tensor $g_{\alpha\beta}(\psi)$  is a  space-time metric tensor in  PFRs. 

Any gauge transformation $\psi_{\alpha \beta} \rightarrow \overline{\psi}_{\alpha \beta}$ 
of the unknown tensor field $\psi_{\alpha \beta}$  induces some mapping  
$g_{\alpha\beta} \rightarrow \overline{g}_{\alpha \beta}$  of Riemannian space-time which are not related with
coordinate transformations.  To ensure the invariance of the equations of the motion of test particles with respect to such
mappings, correct
differential equations for finding tensor field $g_{\alpha \beta}$ must be invariant with respect to geodesic mappings
of Rimmanian's space-time.  \citep{verozub91}.
These transformation  play role of gauge transformation in the theory under consideration.
 The simplest geodesic-invariant generalisation of the vacuum Einstein equations  are:
\begin{equation}
B_{\alpha\beta;\gamma}^{\gamma}-B_{\alpha\delta}^{\epsilon}B_{\beta\epsilon
}^{\delta}=0.\label{myeqs}%
\end{equation}

These  equations are vacuum bimetric differential equations for the tensor
\begin{equation}
B_{\alpha\beta}^{\gamma}=\Pi_{\alpha\beta}^{\gamma}-\overset{\circ}{\Pi
}_{\alpha\beta}^{\gamma}.\label{tensB}%
\end{equation}
(Greek indices run from 0 to 3), where
\begin{equation}
\Pi_{\alpha\beta}^{\gamma}=\Gamma_{\alpha\beta}^{\gamma}-(n+1)^{-1}\left[
\delta_{\alpha}^{\gamma}\Gamma_{\epsilon\beta}^{\epsilon}+\delta_{\beta
}^{\gamma}\Gamma_{\epsilon\alpha}^{\epsilon}\right]  ,\label{Thomases}%
\end{equation}%
\begin{equation}
\overset{\circ}{\Pi}_{\alpha\beta}^{\gamma}=\overset{\circ}{\Gamma}%
_{\alpha\beta}^{\gamma}-(n+1)^{-1}\left[  \delta_{\alpha}^{\gamma}%
\overset{\circ}{\Gamma}_{\epsilon\beta}^{\epsilon}+\delta_{\beta}^{\gamma
}\overset{\circ}{\Gamma}_{\epsilon\alpha}^{\epsilon}\right]
,\label{Thomases0}%
\end{equation}
$\overset{\circ}{\Gamma}_{\alpha\beta}^{\gamma}$ are the Christoffel symbols
of the Minkowski space-time  , $\Gamma_{\alpha\beta}^{\gamma}$ are the Christoffel
symbols of the Riemannian space-time, whose
fundamental tensor is $g_{\alpha\beta}$. The semi-colon 
denotes the covariant differentiation in  Minkowski space-time.

Some additional conditions
can be imposed on the tensor $g_{\alpha\beta}$  because of the geodesic  invariance of the equations.
 In particular,
under the conditions  $Q_{\alpha}=\Gamma_{\alpha\sigma}^{\sigma}-\overset{\circ}{\Gamma}%
_{\alpha\sigma}^{\sigma}=0 \label{gaugeConditions}$
Eqs. (\ref{myeqs}) are reduced to  Einstein's  vacuum
equations .  Therefore, for solving  of  concrete problems  it is sufficient to solve the system of the differential equations:
\begin{equation}
\label{Vr_Einstein Eqs}
R_{\alpha\beta}=0; \, \, 
Q_{\alpha}=0 
\end{equation}
in a selected coordinate system of Minkowski space-time
by using mathematical methods elaborated for  General Relativity. 
The additional conditions are covariant and do  not impose any restrictions  on   the coordinate system selection.

The spherically- symmetric solution of these
equations in  Minkowski space-time   have no an event
horizon  and  physical singularity in the centre \citep{verozub91, verozub96}.
If the distance from a point  attractive mass is many larger than $  r_{g}$, 
the
physical consequences, resulting from these equations, are very close to  General Relativity results.
However, they are principally  other at  short distances from the central mass .
 
\section{Supermassive configurations}

It follows from Eqs. (\ref{myeqs})  that the gravitational force of a point mass $M$ affecting a
free-falling particle  of mass $m$ is given by \citep{verozub91,  verozub96}.
\begin{equation}
F=-m\left[c^{2} C^{\prime}/2A +(A^{\prime}/2A  - 2 C^{\prime}/2C)  \overset{\cdot}{r}^{2}\right]  ,
\label{gravaccel1}%
\end{equation}
where $A=f^{\prime2}/C,\ C=1-r_{g}/f,\ f=(r_{g}^{3}+r^{3})^{1/3}$, 
$r_{g} =2GM/c^{2}$, $r$ is the radial distance from the centre, 
 $G$ is the
gravitational constant, $c$ is the  speed of light at infinity, the prime denotes 
 derivatives with respect to $r$, $\dot{r}=dr/dt$. 

 For   particles at rest ($\overset{\cdot}{r} =0$ ) 
\begin{equation}
F=-\frac{GmM}{r^{2}} \left[
1-\frac{r_{g}}{(r^{3}+r_{g}^{3})^{1/3}} \right]
\label{ForceStat}%
\end{equation}

\begin{figure}
\centering
\includegraphics[width=0.8\linewidth]{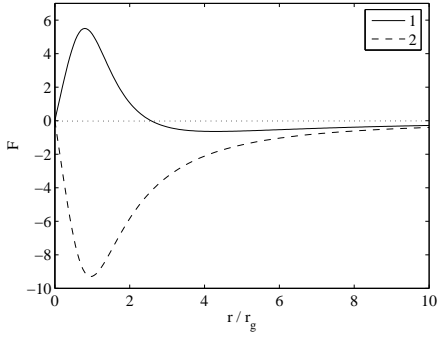}
 \caption{The gravitational force (arbitrary units) affecting  free-falling  particles  (curve
1) and   particles  at rest (curve 2) near a point  attractive
mass $M$.}
 \label{GRForce}
\end{figure}

 Fig. \ref{GRForce} shows the force $F$ 
affecting    particles at rest  and  particles, free
falling from infinity with zero initial speed,   as a
function of the distance $\overline{r}=r/r_{g}$ from the centre.
It follows from  this figure  that the gravitational force affecting a particle at rest  tends to zero when distance from the centre $r\rightarrow 0$.  It would therefore be interesting to know which masses of the equilibrium
configurations can exist if the gravitational force is given by Eq.
(\ref{ForceStat} ). To answer this question one  can  proceed from the equation of
the hydrostatic equilibrium%
\begin{eqnarray}
\label{EquilibrHydEq}
dp_{m}/dr=\rho_{m} g_{m},  \,\, dM_{r} /dr=4 \pi r^2 \rho_{m},\, \,   \\
 p_{m}=p_{m}(\rho_{m}).  \nonumber
\end{eqnarray} 
 In these equations  $M_\mathrm{r}$ is the  mass  of matter  inside  the sphere $\mathcal{O}_\mathrm{r}$  of the radius $r$, $g_{m}=F/m$  is the force   (\ref{ForceStat}) per unit of mass,  $r_{g}=2 G \, M_{r}  /c^2$ is the Schwarzschild radius of  the sphere $\mathcal{O}_{r}$  and $p=p(\rho_{m})$ 
 is the equation of state of   matter  inside  the object. We use the equation of state by   \citep{harrison},
  which is 
 valid for  the  density range  of 
   $(8 \div 10^{16} ) \,   {g}   \,  {cm}^{-3}$.

  For the  gravitational force under consideration there are 
 two types of the solutions of  Eqs. ( \ref{EquilibrHydEq} ).  Besides the solutions describing white dwarfs and newtron stars,
  there are     solutions  with very large masses. For example, for the mass of $ 3\,10^{6} M_{\odot}$ 
  the object radius resulting from Eqs.  (\ref{EquilibrHydEq} )   is about
 $0.4 R_{ \odot}=0.034\,  r_{g}=3 \,10^{10}{cm}$  where $r_{g}$ is the Schwarzschild radius  of the object \citep{verozub96}.  These configurations are stable \citep{verkochStab}.

\section{A Supermassive Object in the Galaxy Centre}

The rate of  gas accretion onto
the supermassive object in the Galaxy centre,   associated with radio-source Sgr A* ,  due to  star winds 
from  surrounding young stars is of the order of $10^{-7}$
$M_{\odot}\ yr^{-1}$.   Therefore, if the object has a
solid surface, it can have   an  atmosphere, basically  hydrogen. 
Estimates show that 
in a  time of $10$ $Myr$ (an estimated lifetime of  surrounding
stars) the mass $M_{atm}$ of the gaseous envelope can  reach
$10 \, M_{\odot}$. 
  At the temperature
$T=10^{7} K $  the atmosphere height
$h_a \sim  10^{8}{cm}$, and the maximal atmospheric density
$\rho_a  \sim 10^{3}\,   {g} \,{cm}^{-3}$.
Under such a condition,  a  hydrogen combustion 
 must begin already in our time. 
Calculations show \citep{verozub06} that at 
$T=10^{7} K $  the luminosity of the object due to the  nuclear bearing on  the surface
$L=2.4\, 10^{31}{erg}\ {s}^{-1}$
at $\rho_{a}=10^{2}{g} \,{cm}^{-3}$ and $L=5.6 \, 10^{33}{erg}\, {s}^{-1}$ at $\rho_{a}%
=10^{3}{g} \, {cm}^{-3}$. 


Due to   gravitational redshift a local frequency (as  measured by a local observer)    differs  from the one for a remote observer by the factor  $1+z_{g}=1/ \sqrt{C}$
where the function  $C=C(r)$   is defined in Eqs. (\ref{gravaccel1}).  
The difference between   local and observed frequency can be significance for the  emission emerging from small distances from the surface.  
For example, at  the above luminosity $L=5.6\, 10^{33}\,  {erg} \, {s}^{-1}$ 
the local frequency of the maximum of the blackbody 
  emission is 
equal to $5.7\, 10^{14} Hz$. The observed frequency of the maximum  is equal to $2.7 \, 10^{12} Hz$.   The corresponding specific luminosity  is   
 $ \sim   2\, 10^{21}\, {erg} \, {s}^{-1} \mathrm{Hz} ^{-1}$.
At  $L=10^{36}\,  {erg} \, {s}^{-1}$ 
the frequency of the maximum is equal to $2 \,10^{15} \,  Hz $ ,
the observed frequency is equal to $9.7 \,10^{12} \,  Hz $, and 
the specific luminosity is  $ \sim   1\, 10^{23}\, {erg} \, {s}^{-1} Hz ^{-1}$.
These magnitudes are close to observation data    \citep{melia}.

\begin{figure}[h]
\centering
\includegraphics[width=0.8\linewidth]{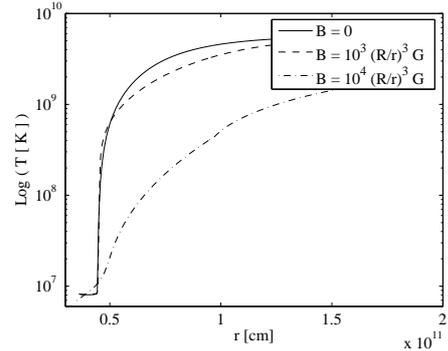}
 \caption{A tipical  temperature    jump as the function of  the distance 
 $r$ from the centre inside of the region  of protons stopping for the object of the radius
$R=0.04 r_{g}$.
 The plots for three value of the magnetic field $B$ are shown.
 } 
 \label{JumpT}
\end{figure}

A hot layer appears in the  atmosphere  because of   deceleration and stopping of   infalling protons. 
Fig. \ref{JumpT} shows a typical temperature profile in the region of  the infalling protons stopping.

\begin{figure}[tbh]
\centering
\includegraphics[width=0.8\linewidth]{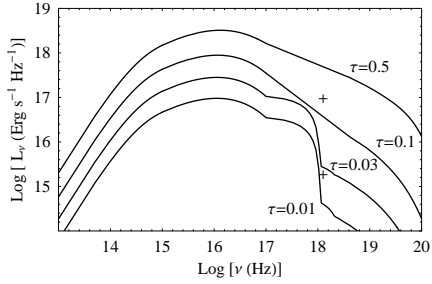}
 \caption{ Emergent spectra of    nuclear burning after comptonization by a  homogeneous     hot layer for
 the optical thickness $\tau=0.01,\, 0.03\, ,0.1,\, 0,5$ . The crosses denote the observation data  \citep{baganoff}, 
 \citep{ porquet}.}
  \label{fig: ComptonSpectrum}
\end{figure}

Because of comptonization of  the Plancian radiation of the burning layer on the surface of the central object this radiation  can
manifest itself  at  high frequencies. 
Fig \ref{fig: ComptonSpectrum} shows 
  an approximate emergent spectrum
of the nuclear radiation  at the surface  after
comptonization   obtained for the case when
  the  temperature at the surface is $10^7 K$ and $\rho_{a}=3\, 10^{3} {g}{cm}^{-3}$.
   The results are plotted  for several values of the optical thickness $\tau$ .

It follows from this  figure that   the
spectrum  may  give a contribution to the observed Sgr A* radiation
\citep{baganoff, porquet}  in the range $10^{15} \div 10^{20}\, Hz$ .

It is necessary to note that the
 timescale $\delta t$ of a variable process is connected with the size $\Delta $
 of its region by the relationship $ \Delta \leq c\ \delta t/(1+z_{g})$.
 For example,  the variations
 in the radiation intensity of $\sim 600\ s$ on  clock of  a remote observer occur    at the
distance $d \leqslant 1.7 R$ from the surface. For this reason high-energy flashes in the Galaxy centre 
   can be
interpreted as processes   near the surface of the central objects.

\begin{figure}[tbh]
\centering\includegraphics[width=0.8\linewidth]{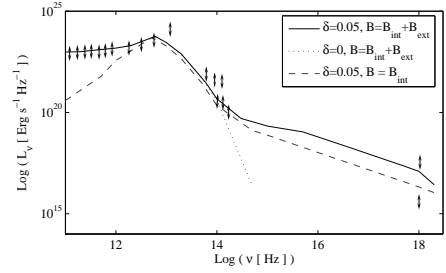}
\caption{The spectrum of the synchrotron radiation}
 \label{spectrumS}
\end{figure}

Some authors \citep{robertson}  find in spectra of
candidates into  Black Holes some evidences for the existence of  a  proper magnetic
field of these objects that is incompatible with the existence of the 
event horizon. In order to show how  the 
magnetic field near the surface of the object in question can manifests  itself,  we
have found a contribution  of a magnetic field to the spectrum of  the synchrotron
radiation.  We  assume  that the magnetic field   is of the form 
$
B_\mathrm{int}=B_{0}  (R/r)^{3},
$
where $B_{0}=1.5\,10^{4}\  \mathrm{G}$.
Fig. \ref{spectrumS} shows the spectrum of the synchrotron
radiation in the band of $10^{11}\div2\, 10^{18}\ \mathrm{Hz} $ for three
cases:\\
-- for the   magnetic field   $B_\mathrm{int} $   
  of
the object,   supposing  the  gibrid distribution of electrons  \citep{ozel}.  
(the dashed line),\\
-- for   the sum of      $B_\mathrm{int} $   and  an external equipartition magnetic field       $B_\mathrm{ext}=(\dot{M} v_\mathrm{ff}  r^{-2})^{1/2}$           
  which may exist in the
accretion flow (\citep{melia}) 
 ( the solid line),\\
--  for the  magnetic field  $B_\mathrm{int} $ without non-themal electrons  (the dotted line which at the frequencies          $ < 10^{14}\, \mathrm{Hz} $               coincides with the dash line).
 in the presence of  the proper magnetic field 
It follows from the figure  that the proper magnetic field can manifest itself for the remote observer at the frequencies 
larger that $2\cdot 10^{12}\,  Hz $  -- in IR and X-ray radiation.

\section{Acceleration of the  Universe expansion as a consequence of gravity properties}
Consider another consequence of the 
gravitational equations under consideration.
It is follows from fig. \ref{GRForce} that the gravitational force in flat space-time acting on free
moving particles essentially differs from that acting on the particles at
rest.

\begin{figure}[h]
\centering
\includegraphics[width=0.8\linewidth] {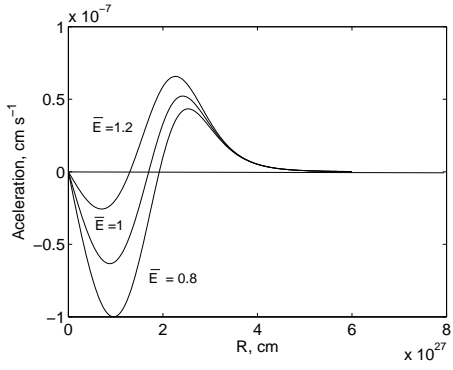}
\caption{The acceleration of particles on the surface of an sphere of the
radius $R$ for three value of the parameter $\overline{E}$. The matter
density is equal to $10^{-29}{g}\,{cm}^{-3}$.}
\label{fig:aceleration}
\end{figure}

A magnitude which is related with  observations in the expanding Universe
is a relative velocity of distant star objects with respect to an observer.
The radial velocity $v=\dot{R}=dR/dt$ of particles on the surface of a selfgravitating expanding 
homogeneous sphere of a radius $R$ 
 can be obtained from equations of the motion of a test particle
 \citep{verozub91}:
\begin{equation}  \label{StarVelocity}
v=c\frac{C f^{2}}{R^{2}}\sqrt{1-\frac{C}{\overline{E}^{2}}},
\end{equation}
where $A$ and $C$ are the functions of the distance $R$ of the object from an
observer in which  $M=(4/3) \pi \rho R^{3}$ is the matter mass inside of the sphere
of the radius $R$,  $\rho$ is the observed matter density, and $r_{g}=(8/3)\pi c^{-2}G\rho R^{3}$ is 
Schwarzshild's radius of the matter inside of the sphere. The parameter
  $\overline{E}$ is the total energy of a  particle divided by $m c^{2}$.
Fig. \ref{fig:aceleration} shows the radial acceleration $\ddot{R}=v'\dot{v}$ of 
a particle on the surface of the sphere of the radius $R$ in flat space-time 
Two conclusions can be made from this figure.\newline
\noindent 1. At some distance from the observer the relative acceleration
changes its sign. If the $R < 2 \cdot 10^{27}{cm}$, the radial
acceleration of particles  is negative. If $R > 2 \cdot 10^{27}{cm}$,  the 
acceleration is positive. Hence, for
sufficiently large distances the  gravitational force affecting  particles 
is repulsive and gives rise to a relative acceleration of   particles. \\
2. The gravitational force, affecting  the particles, tends to zero when $R$
tends to infinity.  (The same fact takes place as regards the force acting
on particles in the case of a static matter .)  The reason of the fact is that
at a sufficiently large distance $R$  from the observer the Schwarzschild radius of
the matter inside the sphere of the radius $R$ become of the same order as its radius.
Approximately at $R \sim 2\, r_{g}$ the gravitational force begin to decrease.
The ratio $R/r_{g}$ tends to zero when $R$ tends to infinity, and under the
circumstances the gravitational force in the theory under consideration
tends to zero. 

The above ball can be consedered as a part of a flat  accelerating  Universe.
To calculate the velocity of a particle at the distance $R$ from the observer there is no necessity to demand a global
spherical symmetry of matter outside of the sphere  due to  the second properties of the gravitational force,  
because  the gravitational influence of very distant matter is neglected small. Therefore a relative 
velocity of   particles at the distance $R$ from the observer is determined by  gravitational field of the matter
inside of the sphere of the radius $R$. 

Proceeding from this equation we can  find Hubble's diagram,  following basically  to  the method being used in 
  \citep{zeldovich}.
 Let $\nu_{0}$
be a  local frequency in the proper reference frame of a moving source at the distance $R$ from an
observer, $\nu_{l}$ be this frequency in a local
inertial frame, and $\nu$ be the frequency as measured by the observer.in the
sphere centre. The redshift $z=(\nu_{0}-\nu)/ \nu$ is caused by both 
Doppler-effect and gravitational field. The Doppler-effect is a consequence
of a difference between the local frequency of the source in inertial and
comoving reference frames, and it is given by  $\nu_{l}=\nu_{0}\, [1-\sqrt{(1-v/c)(1+v/c) } ] $.

The gravitational redshift is caused by the matter inside of the sphere of the
radius $R$.  
The local frequency  $\nu_{0}$ at the distance $R$
from an observer are related with  the observed frequency  $\nu$ by equality
$\nu= \nu_{l} \sqrt{C}$. 
 Thus,  the relationship between
frequency $\nu$ as  measured by the observer and the proper frequency $\nu_{0}$ 
of the moving source in the gravitational field takes the form 
\begin{equation}  \label{TotalRedshift}
\nu/ \nu_{0} =\left[  C \, (1-v/c)/(1+v/c)\right]^{1/2}.
\end{equation}

Equation (\ref{TotalRedshift}) 
yields the quantity $z=\nu_{0}/\nu-1$  as a function of $R$. By
solving this equation numerically  we obtain the
dependence $R=R(z)$   of the  measured distance  $R$   as a function of the redshift.
Therefore the distance modulus to a distant star object is given by 
\begin{equation}  \label{muofz}
\mu=5\,  log_{10}[R(z)\, (z+1)] -5
\end{equation}
where $R(1+z)$ is a bolometric distance (in $pc$) to the object.

If we demand that  equation  (\ref{StarVelocity}) must to give a correct radial velocity of 
distant star objects  in the expansive Universe, it must to  lead to the Hubble
law at small distances R. At this condition the Schwarzschild radius $%
r_{g}=(8/3) \pi G \rho R^{3}$ of the matter inside of the sphere is very small
compared with $R$. For this reason $f\approx r$, and $C=1-r_{g}/r$. Therefore, at 
$\overline{E}=1$,  and we obtain from (\ref{StarVelocity}) that 
$v=H R$ where $H=\sqrt{(8/3) \pi G \rho}.$
If $\overline{E}\neq 1$ equation (\ref{StarVelocity}) does not lead to the
Hubble law,  since $v$ does not tend to $0$ when $R\rightarrow 0$. For this reason we set 
$\overline{E}= 1$ and look for a  value of the density at which a good
accordance with observation data can be obtained.

\begin{figure}[htb ]
\centering
 \includegraphics[width=0.8\linewidth]{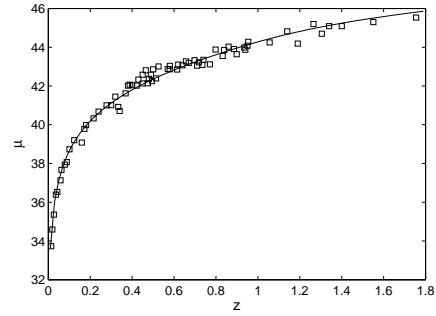}
\caption{The distance modulus $\protect\mu$ vs. the redshift $z$ for the density 
$\protect\rho=4.5 \cdot 10^{-30} g\, cm^{-3}$. Small squares denote the
observation data according to Riess et al. }
\label{fig:HubbleFig}
\end{figure}

The fig. (\ref{fig:HubbleFig}) show the Hubble diagram  compared with observations data  \citep{riess} . 
It follows from this figure that the model under consideration does not
contradict observation data. 

With the value of the density $\rho=4.5 \cdot
10^{-30} g\, cm^{-3}$ we obtain  that  $H=1.59\cdot 10^{-18} c^{-1}=49 km/c^{-}\, Mpc.$

\section{Testing by PSR 1913+16}

Our   generalisation of the Einstein equations
with a matter source  are of the form \citep{verkochPuls}
\begin{equation}
B_{\alpha\beta;\gamma}^{\gamma}-B_{\alpha\sigma}^{\gamma}B_{\beta\gamma
}^{\sigma}=k\left(  T_{\alpha\beta}-1/2G_{\alpha\beta}T\right)
,\label{MainEquationWithSource}%
\end{equation}
where $k=8\pi G/c^{4}$, $T_{\alpha\beta}$ is the matter energy-momentum
tensor, $T=G^{\alpha\beta}T_{\alpha\beta}$, and
\begin{equation}
G_{\alpha\beta}=g_{\alpha\beta}-(n+1)^{-1}Q_{\alpha}Q_{\beta}%
,\label{5dimMetricTensorInvatiant}%
\end{equation}

These  equations with a matter source   were tested \citep{verkochPuls} by binary pulsar  PSR 1913+16
in the framework of the   Blandford - Teukolsky model \citep{blandford}.
Namely, we take into account the dependence of the phase of the
pulsar radiation on the following tree  observable parameters
which are  functions of the
unknown masses $m_{1}$ and $m_{2}$ of two pulsars: 
1)  The relative rate  of the  orbital period $P_{b}$ increasing:  $dP_{b}/dt=\overset{\cdot}{P}
_{b}=\overset{\cdot}{P}_{b}(m_{1,}m_{2})$, 2) The rate of periastron shift of the pulsar orbit  $\overset{\cdot}{\omega}
=\overset{\cdot}{\omega}(m_{1,}m_{2})$, 3) the quantity  $ \gamma=(t_{1}-t_{2})/{\sin(\widetilde{E})}$
where $t_{1}$ is the moment of the radiation of a pulse as measured by
a remote observer, $t_{2}$ is the same
moment as measured in the proper frame of reference of the pulsar,
$\widetilde{E}$ is the eccentric anomaly.

For  a system of two gravitationally bound point masses $m_{1}$ and
$m_{2}$   we have obtained the equation
\begin{equation}
\frac{dP_{b}/dt}{P_{b}}=\frac{92\pi}{5c^{5}}\>\left(  2\pi G\right)
^{5/3}P_{b}^{8/3}\>f(e)m_{1}m_{2}M_{p}^{-1/3}\;, \label{dP/dt}%
\end{equation}
where  $M_{p}=m_{1}+m_{2}$,
\begin{eqnarray}
f(e)=\left(  1-e^{2}\right)  ^{-7/2}\left(  1+73/24\>e^{2}
+37/96\>e^{4}\right)   
\end{eqnarray}
and $e$ is the eccentricity 
The rate of the periastron shift of the  pulsar orbit resulting 
 from our gravitation equations is given by 
\begin{equation}
\dot{\omega}=\frac{6\pi GM}{a\left(  1-e^{2}\right)  c^{2}}+\frac{\pi
G^{2}M^{2}}{2a^{2}\>\left(  1-e^{2}\right)  ^{2}\>c^{4}}\>f_{1}\>(e) ,
\label{domega}%
\end{equation}
\citep{verozub96}
where $\ f_{1}\>(e)=\left(  54+16e^{2}-e^{4}\right)  $. The second term for
the considered system is about $0.01$ of the total value of $\dot{\omega}$
and, therefore must be taken into account. 

 For the value of $\gamma$ we have obtained 
 
\begin{equation}
\gamma=\frac{GP_{b}em_{2}^{2}(m_{1}+2m_{2})}{2\pi a c^{2}(m_{1}+m_{2})^{2}} .
\label{ParameterGammaPulsar}%
\end{equation}

\begin{figure}[tbh]
\centering
\includegraphics[width=0.8 \linewidth]{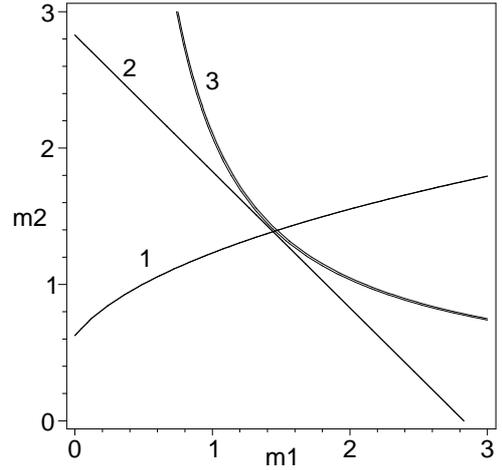}
\caption{he plots of the functions $\gamma=\gamma(m_{1},m_{2})$ (curve 1),
$\omega=\omega(m_{1},m_{2})$ (curve 2), $\dot P_{b}=\dot P_{b}(m_{1},m_{2})$
(curve 3).}
\label{pulsar}
\end{figure}

Fig. \ref{pulsar} shows that in the plane $m_{1}m_{2}$ the 
curves $\overset{\cdot}{P}
_{b}=\overset{\cdot}{P}_{b}(m_{1,}m_{2})$,
$\overset{\cdot}{\omega} =\overset{\cdot}{\omega}(m_{1,}m_{2})$ and 
$\gamma=\gamma(m_{1}m_{2})$ meet in one point.
The value of the total  mass $M_{p}$ of the system resulting from the measured value of  $P{_b}$ and $e$ 
by equation (\ref{domega}) is equal to
$(2.82845\pm0.03) M_{\odot}$. But then, by using  (\ref{ParameterGammaPulsar}) one can find that the pulsar masses 
 are equal to $(1.441\pm0.03) M_{\odot}$ and $(1.387\pm0.03) M_{\odot}$. These results differ very little from
those in General Relativity \citep{taylor}:  $(1.442\pm0.03) M_{\odot}$ and $(1.386\pm0.03) M_{\odot}$, correspondingly. Due to
a kinematic effect in our  Galaxy \citep{damour} a little correction $(-0.017\pm 0.005)\cdot 10^{-12}$ must  be added to the found theoretical value of $\dot{P}_{b} =(-2.40249\pm 0.005)\cdot 10^{-12} $  Taking into account this correction,
the ratio of the observed value of  $\dot{P}_{b}$ to that found theoretically is equal to $1.0023\pm0.0047$.

\end{document}